\documentclass[preprint,prl]{revtex4}

\usepackage[final]{graphicx}
\usepackage{dcolumn}
\usepackage{bm}

\begin{document}

%
%
%

\title{Cavity QED System with Optical Lattice:Photon Statistics and Conditioned Homodyne Detection}


\author{Mambwe Mumba}%
\affiliation{Intel Corporation, 2200 Mission College Blvd, Santa Clara, CA 95054}
\author{Dyan Jones}
\affiliation{Department of Physics, Mercyhurst University, Erie PA 16546}
\author{Perry Rice}
\affiliation{ Macklin Quantum Information Sciences, \\ Department of Physics, Miami University, Oxford,
Ohio 45056}%
 \email{perryr@uoregon.edu}

\date{\today}

\begin{abstract}
We investigate the quantum fluctuations of a single atom in a
weakly driven cavity, with an intracavity optical lattice. The weak driving field is on
resonance with the atoms and the cavity, and is the second-harmonic of the lattice beam.
In this special case we can find eigenstates of the Hamiltonian in terms of Mathieu functions. We present analytic results for the
second order intensity correlation function $g^{(2)}(\tau)$ and
the intensity-field correlation function $h_{\theta}(\tau)$, for
both transmitted and fluorescent light for weak driving fields. We
find that the coupling of the center of mass motion to the
intracavity field mode can be deleterious to nonclassical effects
in photon statistics; less so for the intensity-field correlations, and compare
the use of trapped atoms in a cavity to atomic beams.

\end{abstract}

  \maketitle
\section{Introduction}

 One system that has long been a
paradigm of the quantum optics community is a single-atom coupled
to a single mode of the electromagnetic field, the Jaynes-Cummings
model\cite{JC}. In practice the creation of a preferred
field mode is accomplished by the use of an optical resonator.
This resonator generally has losses associated with it, and the
atom is coupled to vacuum modes out the side of the cavity leading
to spontaneous emission. Energy is put into the system by a
driving field incident on one of the end mirrors. The
investigation of such a system defines the subfield of cavity
quantum electrodynamics\cite{CQED}.  Cavity QED systems exhibit many of the
nonclassical effects described above, as well as interesting
nonlinear dynamics which can lead to optical
bistability\cite{bistable}, or chaotic dynamics\cite{chaos}. The
presence of the cavity can also be used to enhance or reduce the
atomic spontaneous emission rate\cite{CQED}. This system has also
been studied extensively in the laboratory, but several practical
problems arise.\cite{expt1,expt2,thy} There are typically many atoms in the cavity at
any instant in time, but methods have been developed to load a
cavity with a single atom. A major problem in experimental cavity
QED stems from the fact that the atom(s)are not stationary as is
often assumed by theorists. The atoms have typically been in an
atomic beam originating from an oven, or perhaps released from a
magneto-optical trap. This results in inhomogeneous broadening of
the atomic resonance from Doppler and/or transit-time broadening.
Using slow atoms can reduce these effects, but the coupling of the
atom to the light field in the cavity is spatially dependent, and
as the atoms are in motion, the coupling is then time dependent;
also different atoms see different coupling strengths.

With greater control in recent years of the center of mass motion
of atoms, developed by the cooling and trapping community,
preliminary attempts have been made to investigate atoms trapped
inside the optical cavity\cite{Cool}. The recent demonstration of
a single atom laser is indicative of the state of the art
\cite{HJKSAL}. In this paper we consider a
single atom cavity QED system with the addition of an external
potential, provided perhaps by an optical lattice, and study the
photon statistics and conditioned field measurements of both the
transmitted and fluorescent fields. We seek to understand (with a
simple model at first) how the coupling of the atom's center of
mass motion to the light field affects the nonclassical effects
predicted and observed for a stationary atom.

The system we consider is shown schematically in Fig. 1.
\begin{figure}
   \begin{center}
   \begin{tabular}{c}
   \includegraphics[height=6cm]{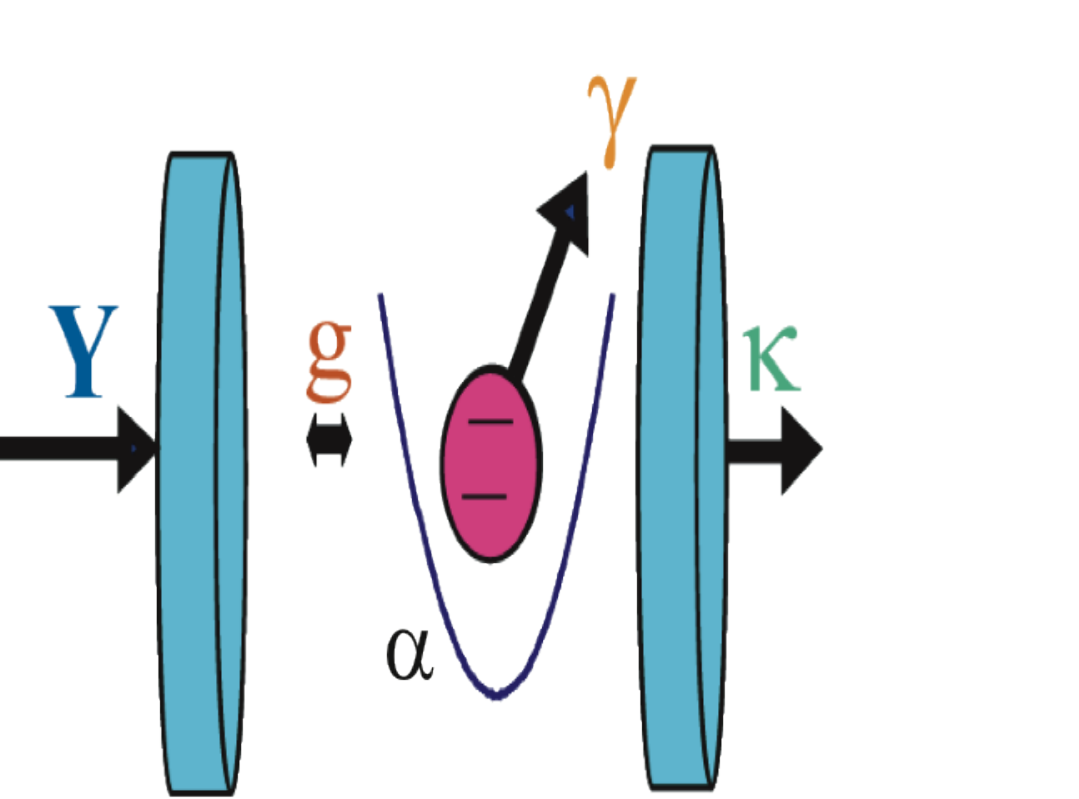}
   \end{tabular}
   \end{center}
   \caption[example]
{Single atom in a weakly driven optical cavity. Here g is the reversible coupling rate between the mode of the cavity and the atom, $\kappa$ is the decay rate of the field mode of the cavity, $\gamma$ is the spontaneous emission rate. $Y$ is the external drive (taken to be a classical field).}
   \end{figure}

\begin{equation}
H = \frac{p^2}{2m} + \frac{1}{2}V_0 \cos^2{kz} + \hbar
g_m\cos{k_L}\left(a^\dagger\sigma_+ + a\sigma_-\right).
\end{equation}

A simplifying assumption is that $k=2k_L$ which is easily
recreated in the lab through the use of a $\chi^{(2)}$
non-linearity so that $z = k_L x$.  This then reduces the
Schrodinger equation to

\begin{equation}
\frac{d^2\psi}{dx^2} + \frac{2m}{\hbar^2} V_0 \cos{2k_L x} +
g_0\cos{kx}\sqrt{n}\psi = -\frac{2mE}{\hbar^2}
\end{equation}

where we have taken advantage of the fact that by working in the
dressed-state picture, the Jaynes-Cummings term can be substituted
with the eigenvalue $\sqrt{n}$.

Defining three constants:

\begin{eqnarray}
z = k_L x \\
a = \frac{2mE}{k_L^2 \hbar^2}\\
q = \frac{2m}{k_L^2 \hbar^2}\left( V_0 \pm g_0\sqrt{n}\right)
\end{eqnarray}

so that the Schrodinger equation can be written in a form that
looks conveniently like the general form for the Mathieu functions
as described in section 4.2 leads to

\begin{equation}
\frac{d^2 \psi}{dz^2} + \left(a - 2q\cos{2k_Lz}\right)\psi = 0
\end{equation}


The next step is to look at the probability that a transition
between the ground and excited states of an atom will occur. In
order to do this, we must examine how the transitions depend on
the vibrational modes for the different electronic configurations.
When a spatial overlap exists between the vibronic states (refer
to figure \ref{FranckCondonNew.pdf}), the greater transition rates occur
for the larger overlaps.  This \lq\lq overlap\rq\rq\ is described
by using what are known as the Franck-Condon Factors.  Note that
though we talk about the spatial overlap between transitions,
there is no true spatial displacement in either the simple
harmonic or Mathieu cases, so that the Franck-Condon factors are
modelling the atomic coupling to the field lattice.

\begin{figure}
   \begin{center}
   \begin{tabular}{c}
   \includegraphics[height=6cm]{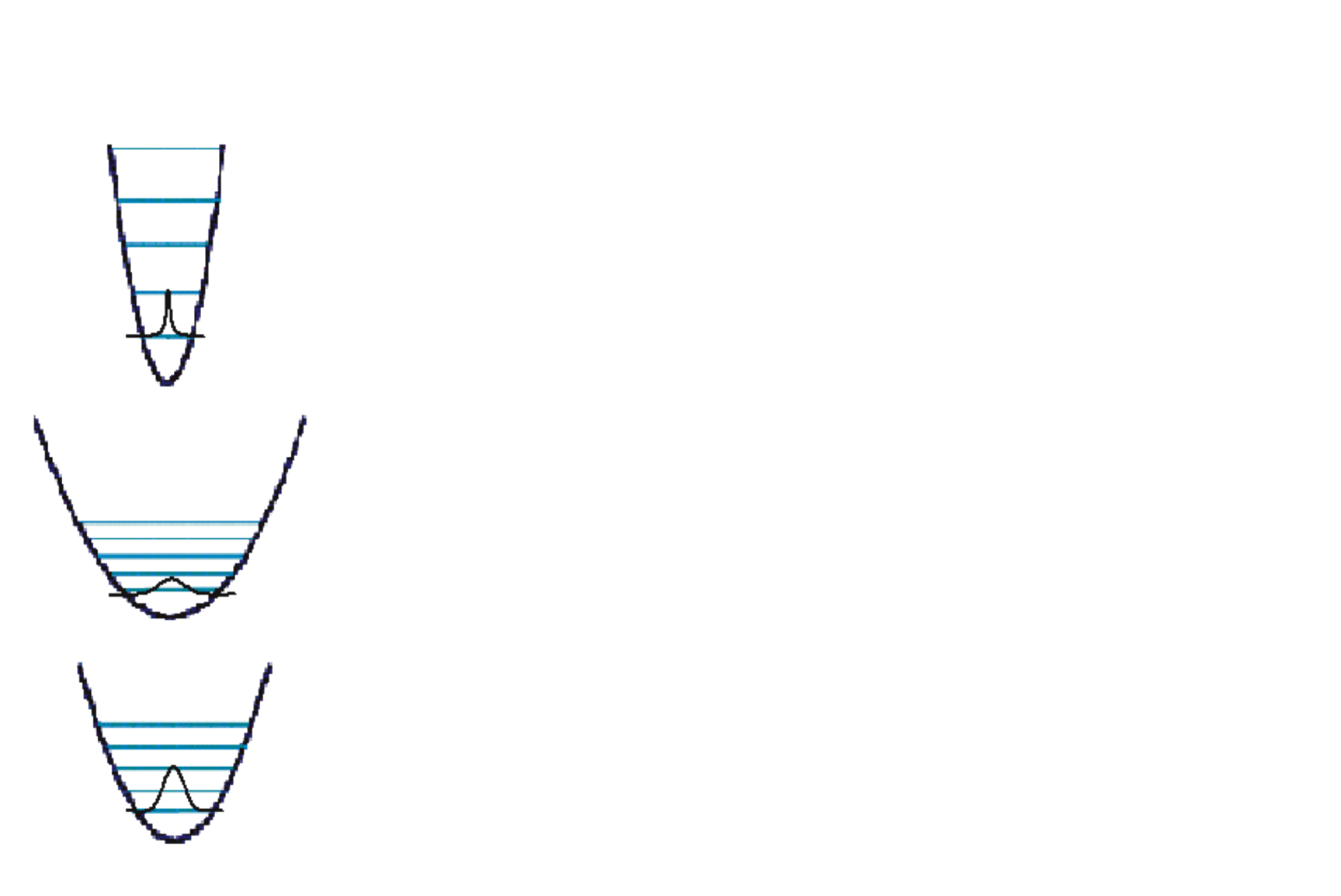}
   \end{tabular}
   \end{center}
   \caption[example]{Need for Franck Condon factors}
\end{figure}

It has already been shown that the coupling rate is given as
\begin{equation} g=\mu_{eg}\sqrt{\frac{\hbar\omega_o}{2\epsilon_0
v}}\cos{kz}
\end{equation}

where
\begin{equation}
\mu_{eg}=\langle e|qz|g\rangle
\end{equation}

However, for completeness the Franck-Condon factors need to be
included, which will now lead to

\begin{equation}
\mu_{eg}=\langle e|qz|g\rangle\langle k_o|k_\pm \rangle
\end{equation}

Our goal is then to express those Franck-Condon factors in terms
of the dressed states of the harmonic oscillator which we have
been dealing with:

\begin{equation}\langle k_o|k_\pm \rangle =
\int \psi_{k_0}^{*} \psi_{k_\pm}dy
\end{equation}

In the harmonic limit, both $\psi_{k,0}$ and $\psi_{k,\pm}$ can be
expressed as a Gaussian function times a Hermite polynomial (for a
full explanation, see \cite{Mambwe}).  However, in terms of the
Mathieu functions, this needs to be altered slightly such that

\begin{equation}
\langle k_0 | k_\pm \rangle = \int \psi^*_{n,k,\pm}\psi_{n,k,0}dy,
\end{equation}

Where some combination of the Ce and Se Mathieu wave functions  needs to be included.  This
needs to be done analytically, so it will not be necessary to take
this any further.  The analytical solution involves using the Ce
and Se eigenfunctions to solve for the value of q once a potential
has been specified.
the equation for the $\dot{C}$'s, or the probability
amplitudes was derived.  They now need to be written in a slightly
different form in order to hide the time dependence. This is
really a change to a time-dependent basis, and begins by defining
$D$'s in terms of $C$'s such that

\begin{eqnarray}
D_{0,l,g} & = & C_{0,l,g} \\
D_{1,l,\pm} & = & C_{1,l,\pm}e^{-\big(l\Omega_{1,\pm}-l\Omega_{0}
\pm g
\big)it}\\
D_{2,l,\pm} & = & C_{2,l,\pm}e^{-\big( l\Omega_{2,\pm}-l\Omega_{0}
\pm \sqrt{2}g \big)it}
\end{eqnarray}

Because the weak-field limit is being examined, $C_{0,l,g}=1$
which causes $\dot{C}_{0,l,g}=0$. However, the above equations are still in the harmonic
approximation. Going beyond this approximation and again using
Mathieu eigenstates, our $\dot{D}$s become

\begin{eqnarray}
D_{0,l,g} &=& C_{0,l,g} \\
D_{1,l,\pm}&=& C_{1,l,\pm}e^{-i\left( E_{1,l,\pm} - E_{0,l}\right)t/{\hbar}}\\
D_{2,l,\pm} &=& C_{2,l,\pm}e^{-i\left( E_{2,l,\pm} -
E_{0,l}\right)t/{\hbar}}
\end{eqnarray}

Using this, it can be shown that

\begin{eqnarray}
\dot{D}_{0,l,g}  =  0\\
\dot{D}_{1,l,+}  =  -\bigg[ \frac{\gamma}{4}+\frac{\kappa}{2} +i
\big( E_{1,l,\pm} - E_{0,l} + g\big)
\bigg]D_{1,l+}-\frac{Y}{\sqrt{2}}D_{0,l,g}-\bigg[
\frac{\gamma}{4}-\frac{\kappa}{2} \bigg]D_{1,l,-}\\
\dot{D}_{1,l,-} = -\bigg[ \frac{\gamma}{4}+\frac{\kappa}{2} +i
\big( E_{1,l,\pm} - E_{0,l} +g \big)
\bigg]D_{1,l-}-\frac{Y}{\sqrt{2}}D_{0,l,g}-\bigg[
\frac{\gamma}{4}-\frac{\kappa}{2} \bigg]D_{1,l,+}\\
\dot{D}_{2,l,+} = -\bigg[ \frac{\gamma}{4}+\frac{3\kappa}{2} +i
\big( E_{2,l,\pm} - E_{0,l} +\sqrt{2}g \big) \bigg]D_{2,l+}-Y\big(
\frac{1}{\sqrt{2}} + \frac{1}{2}\big)D_{1,l,+} \nonumber\\ +
Y\big( \frac{1}{\sqrt{2}} - \frac{1}{2}\big)D_{1,l,-}-\bigg[
\frac{\gamma}{4}-\frac{\kappa}{2}
\bigg]D_{2,l,-} \\
\dot{D}_{2,l,+} = -\bigg[ \frac{\gamma}{4}+\frac{3\kappa}{2} +i
\big( E_{2,l,\pm}-E_{0,l}-\sqrt{2}g \big) \bigg]D_{2,l+} +
Y\big( \frac{1}{\sqrt{2}} 1 \frac{1}{2}\big)D_{1,l,+} \nonumber\\
- Y\big( \frac{1}{\sqrt{2}} + \frac{1}{2}\big)D_{1,l,-}-\bigg[
\frac{\gamma}{4}-\frac{\kappa}{2} \bigg]D_{2,l,+}
\end{eqnarray}

Now, though, these must be changed to account for the
Franck-Condon factors that were discussed in the previous chapter.

\begin{equation}
g_m \longrightarrow g_{\pm,l,m} = g_m \ast
FC_{\pm,l,m}
\end{equation}

So the above equations are altered to get their final form:

\begin{eqnarray}
\dot{D}_{0,l,g}  =  0\\
\dot{D}_{1,l,+}  =  -\bigg[ \frac{\gamma}{4}+\frac{\kappa}{2} +i
\big( E_{1,l,\pm}-E_{0,l}+g_{1,l,+} \big)
\bigg]D_{1,l+}-\frac{Y}{\sqrt{2}}D_{0,l,g}-\bigg[
\frac{\gamma}{4}-\frac{\kappa}{2} \bigg]D_{1,l,-}\\
\dot{D}_{1,l,-} = -\bigg[ \frac{\gamma}{4}+\frac{\kappa}{2} +i
\big( E_{1,l,\pm}-E_{0,l}+g_{1,l,-} \big)
\bigg]D_{1,l-}-\frac{Y}{\sqrt{2}}D_{0,l,g}-\bigg[
\frac{\gamma}{4}-\frac{\kappa}{2} \bigg]D_{1,l,+}\\
\dot{D}_{2,l,+} = -\bigg[ \frac{\gamma}{4}+\frac{3\kappa}{2} +i
\big( E_{2,l,\pm}-E_{0,l}+\sqrt{2}g_{2,l,+} \big)
\bigg]D_{2,l+}-Y\big( \frac{1}{\sqrt{2}} +
\frac{1}{2}\big)D_{1,l,+} \nonumber\\ + Y\big( \frac{1}{\sqrt{2}}
- \frac{1}{2}\big)D_{1,l,-}-\bigg[
\frac{\gamma}{4}-\frac{\kappa}{2}
\bigg]D_{2,l,-} \\
\dot{D}_{2,l,-} = -\bigg[ \frac{\gamma}{4}+\frac{3\kappa}{2} +i
\big( E_{2,l,\pm}-E_{0,l}-\sqrt{2}g_{2,l,-} \big) \bigg]D_{2,l+} +
Y\big( \frac{1}{\sqrt{2}} 1 \frac{1}{2}\big)D_{1,l,+} \nonumber\\
- Y\big( \frac{1}{\sqrt{2}} + \frac{1}{2}\big)D_{1,l,-}-\bigg[
\frac{\gamma}{4}-\frac{\kappa}{2} \bigg]D_{2,l,+}
\end{eqnarray}

We can also prescribe an initial wave function in terms of Gaussian functions. We can
write

\begin{equation}
\langle \Psi_{CM}|l \rangle
\end{equation}

and consider the wavefunction to be a Gaussian

\begin{equation}
\Psi = Ae^{-y/2\Sigma^2}
\end{equation}

where A is the normalization constant and $\Sigma$ is the width of
the Gaussian.  By defining $\sigma = \Sigma/\sigma_0$ with
$\sigma_0 = \left(\hbar/m\Omega_{0,l}\right)^{1/2}$ and finding
the normalization to be $A = \left(m\Omega_{0,l}/\pi\hbar(2^n
n!)^2 \right)^{1/4}$

then

\begin{equation}
D_{0,l,g} = A\int_{-\infty}^\infty
e^{-y^2/2\sigma^2}e^{-y^2/2}H_l(y)dy
\end{equation}

Releasing the harmonic limit, $\Phi_m$ can be defined as a Mathieu
function of order m, and $G(x)$ as a normalized Gaussian.  Then

\begin{equation}
D_{0,l,g} = \int_{a}^{b} \Phi_l G(x)dx
\end{equation}

where the Gaussian $G(x)$ must be equal to the Error Function such
that

\begin{equation}
\int_{-x}^{x}G(x)dx =
\emph{Erf}\left(\frac{x}{\sqrt{2}\sigma}\right)
\end{equation}

\begin{equation}
\emph{Erf}(x) = 1 - \frac{2}{\sqrt{\pi}}\int_x^\infty e^{-u^2}du
\end{equation}

Recall that when explaining the Quantum Trajectory Formalism in
the previous chapter, the condition of being in the weak field
limit was used.  In the steady state for this case, there is a
very small average photon number, and the probability of getting a
collapse is small as well.  The wavefunction for the steady state
is written as

\begin{equation}
|\Psi_{SS}\rangle = \sum_{n,l}\left( D_{n,l,+}^{ss}|n,l,+\rangle +
D_{n,l,-}^{ss}|n,l,-\rangle \right)
\end{equation}

and the wavefunction after a transmission or fluorescence collapse
as

\begin{eqnarray}
a|\Psi_{ss}\rangle =
\frac{|\Psi_{CT}(0)\rangle}{|\Psi_{CT}(0)\rangle|^2}\\
\sigma_- |\Psi_{ss}\rangle =
\frac{|\Psi_{CF}(0)\rangle}{|\Psi_{CF}(0)\rangle|^2}
\end{eqnarray}

The probability of a transmission (cavity emission) occurring at
$\tau = 0$ is

\begin{equation}
P_T(\tau = 0) = 2\kappa \langle \Psi_{CT}|a^\dagger a
|\Psi_{CT}\rangle
\end{equation}

and similarly for a fluorescence

\begin{equation}
P_F(\tau = 0) = 2\gamma\langle \Psi_{CF}|\sigma_+\sigma_-
|\Psi_{CF}\rangle
\end{equation}

Putting these back into the equation for $g^{(2)}(\tau)$,

\begin{eqnarray}
g^{(2)}_{TT}(\tau) & = & \frac{\langle\Psi_{CT}|a^\dagger
a|\Psi_{CT}\rangle}{\langle\Psi_{ss}|a^\dagger
a|\Psi_{ss}\rangle}\nonumber\\
& = & \frac{\sum_{n,l}n|C_{g,n,1}^{CT}(\tau)|^2}{\sum_{n,l}n|C_{g,n,l}^{ss}(\tau)|^2}\nonumber\\
& = &
\frac{\sum_{l}|C_{g,n,1}^{CT}|^2(\tau)}{\sum_{l}|C_{g,n,l}^{ss}|^2}
\end{eqnarray}

Similarly for the fluorescence,

\begin{equation}
g^{(2)}_{FF}(\tau) =
\frac{\sum_{l}|C_{e,0,l}^{CF}|^2(\tau)}{\sum_{l}|C_{e,0,l}^{ss}|^2}
\end{equation}

In order to define the amplitudes of the states, we need to look
at the wave function at the steady state and also after a
collapse. These are expressed respectively as

\begin{eqnarray}
|\psi_{ss}\rangle = \sum^\infty_{n,l=0} \left(
C_{1,l,+}^{ss}e^{-iE_{1,l,+}t}|1,l,+\rangle +
C_{1,l,-}^{ss}e^{-iE_{1,l,-}t}|1,l,-\rangle\right)\\
|\psi(0)\rangle_{coll} = \sum_{n,l=0}^\infty \left(
C_{g,n,l}^{coll}(t)e^{-iE_{g,n,l}t}|g,n,l\rangle +
C_{e,n,l}^{coll}(t)e^{-iE_{e,n,l}t}|e,n,l\rangle\right)
\end{eqnarray}

where the initial amplitudes of the collapsed states are

\begin{eqnarray}
C_{g,n,l}^{coll}(0) =
\frac{\sqrt{2}C_{g,2,l}^{ss}}{\sum_{n,l}\left( 2|C_{g,2,l}^{ss}|^2
+ |C_{e,1,l}^{ss}|^2 \right)}\\
C_{e,0,l}^{coll}(0) = \frac{C_{e,1,l}^{ss}}{\sum_{n,l}\left(
2|C_{g,2,l}^{ss}|^2 + |C_{e,1,l}^{ss}|^2 \right)}
\end{eqnarray}

Now that all of the foundation has been laid, what exactly is the
probability of getting either a transmission or a fluorescence
event at time $t = \tau$ if a transmission or fluorescence event
occured at time $t=0$?  The four possible combinations are
labelled as TT, FF, TF, or FT, and they are expressed as

\begin{eqnarray}
g^{(2)}_{TT} & = &  \frac{\langle a^\dagger(0)a^\dagger(\tau)a(\tau)a()\rangle}{\langle a^\dagger a \rangle^2}\\
g^{(2)}_{FF} & = &  \frac{\langle \sigma_+(0)\sigma_+(\tau)\sigma_-(\tau)\sigma_-(0)\rangle}{\langle \sigma_+\sigma_-\rangle^2}\\
g^{(2)}_{TF} & = &  \frac{\langle a^\dagger(0)\sigma_+(\tau)\sigma_-(\tau)a(0)\rangle}{\langle a^\dagger a \rangle\langle\sigma_+\sigma_-\rangle}\\
g^{(2)}_{FT} & = &
\frac{\langle\sigma_+(0)a^\dagger(\tau)a(\tau)\sigma_-(0)\rangle}{\langle
a^\dagger a \rangle\langle\sigma_+\sigma_-\rangle}
\end{eqnarray}

The collapse operators have been defined such that a transmission
is $\kappa a$ and a fluorescence is $\sqrt{\gamma}\sigma_-$, so
that our collapsed states are

\begin{eqnarray}
|\psi_c^T \rangle = \frac{a|\psi_{ss}}{|a|\psi_{ss}|^2}\\
|\psi_c^F \rangle =
\frac{\sigma_-|\psi_{ss}}{|\sigma_-|\psi_{ss}|^2}
\end{eqnarray}

and the final form of all our second-order correlation functions
can be shown to be \cite{Joe}

\begin{eqnarray}
g^{(2)}_{TT} = \frac{\sum_{\{m\} = 0}^\infty
|C_{1,g\{m\}}^{CT}(\tau)|^2}{\sum_{\{m\} = 0}^\infty
|C_{1,g\{m\}}^{ss}|^2}\\
g^{(2)}_{FF} = \frac{\sum_{\{m\} = 0}^\infty
|C_{0,e\{m\}}^{CF}(\tau)|^2}{\sum_{\{m\} = 0}^\infty
|C_{0,e\{m\}}^{ss}|^2}\\
g^{(2)}_{TF} = \frac{\sum_{\{m\} = 0}^\infty
|C_{0,e\{m\}}^{CT}(\tau)|^2}{\sum_{\{m\} = 0}^\infty
|C_{0,e\{m\}}^{ss}|^2}\\
g^{(2)}_{FT} = \frac{\sum_{\{m\} = 0}^\infty
|C_{1,g\{m\}}^{CF}(\tau)|^2}{\sum_{\{m\} = 0}^\infty
|C_{1,g\{m\}}^{ss}|^2}
\end{eqnarray}


\subsection{Anti-bunching} \label{Antibunching section}

Once again, the Schwartz inequalities that classical fields obey
are

\begin{eqnarray}
g^{(2)}(0) &\ge& 1\\
g^{(2)}(\tau) &\le& g^{(2)}(0)\\
|g^{(2)}(\tau)-1| &\le& |g^{(2)}(0)-1|
\end{eqnarray}

A violation of the first inequality means that there are no
non-negative probability distributions that describe the field.
The other two inequalities tell us information about the photon
distribution of our source.

The third inequality describes any \lq\lq undershoot" or
 \lq\lq overshoot" properties, represented by \\
 $|g^{(2)}(\tau)-1| >
|g^{(2)}(0)-1|$ and $|g^{(2)}(\tau)-1| < |g^{(2)}(0)-1|$,
respectively.

Most interesting, though, is the second inequality. There are
three possibilities for light: random, bunched, or anti-bunched
(see figure \ref{Bunchingpic}). Random photon sources are
represented by $g^{(2)}(\tau) = 1$, where $g$ is completely
independent of $\tau$.  Bunched light, or Photon-Bunching is
represented by super-Poissonian statistics and the inequality that
$g^{(2)}(0) > g^{(2)}(\tau)$. Lastly, is the case of
sub-Poissonian statistics where $g^{(2)}(0) < g^{(2)}(\tau)$,
known as Anti-Bunching.  In photon anti-bunching, there is a great
probability that photons will be further apart than close
together, making the detection pattern much more uniform.  These
are the states that have no classical description of fields. There
are two ways to describe the amount of anti-bunching of a source.
The first is to use perfect anti-bunching, which is the case where
$g^{(2)}(0)=0$. The more anti-bunched a source is, the closer
$g^{(2)}(0)$ will be to zero. However, anti-bunching may also be
characterized by the slope of $g^{(2)}(\tau)$ from some initial
value of $g^{(2)}(0)$.  The differences
between these two terminologies will be explained later.

In the weak field limit, the field quadrature is given as

\begin{equation}
\langle\hat{a}_\theta\rangle = \sum_l \left(
C^*_{1,l}C_{0,l}e^{-i\theta} + C^*_{0,l}C_{1,l}e^{i\theta} \right)
\end{equation}

and so the correlation function for weak fields is

\begin{equation}
h_\theta(\tau) = \frac{\sum_l \left(
C_{1,l}^{CT*}C_{0,l}^{CT}e^{-i\theta} +
C_{0,l}^{CT*}C_{1,l}^{CT}e^{i\theta} \right)}{\sum_l \left(
C_{1,l}^{ss*}C_{0,l}^{ss} + C_{0,l}^{ss*}C_{1,l}^{ss} \right)}
\end{equation}

Following the same format as when examining the $g^{(2)}$'s, the
four combinations can be written as

\begin{eqnarray}
h_{\theta}^{TT}(\tau) & = & \frac{\langle a_\theta(\tau)\rangle_{CT}}{\langle a_0(\tau)\rangle_{ss}}\\
h_{\theta}^{FF}(\tau) & = & \frac{\langle \sigma_\theta (\tau)\rangle_{CF}}{\langle\sigma_0(\tau)\rangle_{ss}}\\
h_{\theta}^{TF}(\tau) & = & \frac{\langle\sigma_\theta (\tau)\rangle_{CT}}{\langle\sigma_0(\tau)\rangle_{ss}}\\
h_{\theta}^{FT}(\tau) & = & \frac{\langle a_\theta
(\tau)\rangle_{CF}}{\langle\sigma_0(\tau)\rangle_{ss}}
\end{eqnarray}

and expressed in terms of probability amplitudes,

\begin{eqnarray}
h_{\theta}^{TT}(\tau) & = & \frac{\sum_{\{m\}}
C_{1,g,\{m\}}^{CT}(\tau)C_{0,g,\{m\}}^{CT}(\tau)}{\sum_{\{m\}}
C_{1,g,\{m\}}^{ss}C_{0,g,\{m\}}^{ss}} \cos\theta\\
h_{\theta}^{FF}(\tau) & = & \frac{\sum_{\{m\}}
C_{0,e,\{m\}}^{CF}(\tau)C_{0,g,\{m\}}^{CF}(\tau)}{\sum_{\{m\}}
C_{0,e,\{m\}}^{ss}C_{0,g,\{m\}}^{ss}} \cos\theta\\
h_{\theta}^{TF}(\tau) & = & \frac{\sum_{\{m\}}
C_{0,e,\{m\}}^{CT}(\tau)C_{0,g,\{m\}}^{CT}(\tau)}{\sum_{\{m\}}
C_{0,e,\{m\}}^{ss}C_{0,g,\{m\}}^{ss}} \cos\theta\\
h_{\theta}^{FT}(\tau) & = & \frac{\sum_{\{m\}}
C_{1,g,\{m\}}^{CF}(\tau)C_{0,g,\{m\}}^{CF}(\tau)}{\sum_{\{m\}}
C_{1,g,\{m\}}^{ss}C_{0,g,\{m\}}^{ss}} \cos\theta
\end{eqnarray}

{\section{Inequalities and Non-Classical Behaviors}
\setlength{\parindent}{0.25in}

A set of figures are now presented.  First,
though, a reminder to the reader of the inequalities is included,
and in which case the violations apply.  In the graphs of
$g^{(2)}(\tau)$, the data must be examined in two parts.  The
transmission and fluorescence cases follow a set of inequalities
different from those for the cross correlations.

\subsection{Transmission and Fluorescence}
\setlength{\parindent}{0.25in}

  The inequality satisfied by classical fields with a
positive definite probability distribution is

\begin{equation}
g^{(2)}(\tau)\ge g^{(2)}(0)
\end{equation}

Therefore, violations of this are written as

\begin{eqnarray}
B:     g^{(2)}(0) > g^{(2)}(\tau)\\
A:     g^{(2)}(0) < g^{(2)}(\tau)
\end{eqnarray}

 Simply put, they
are dependent on the initial change in slope of the graph.  An initial decrease in
the graph signifies bunching \lq\lq B \rq\rq, whereas an an
initial increase signifies anti-bunching, \lq\lq A \rq\rq.

\begin{figure}
\centering
         \includegraphics[width=\textwidth,totalheight=2.5in, keepaspectratio]{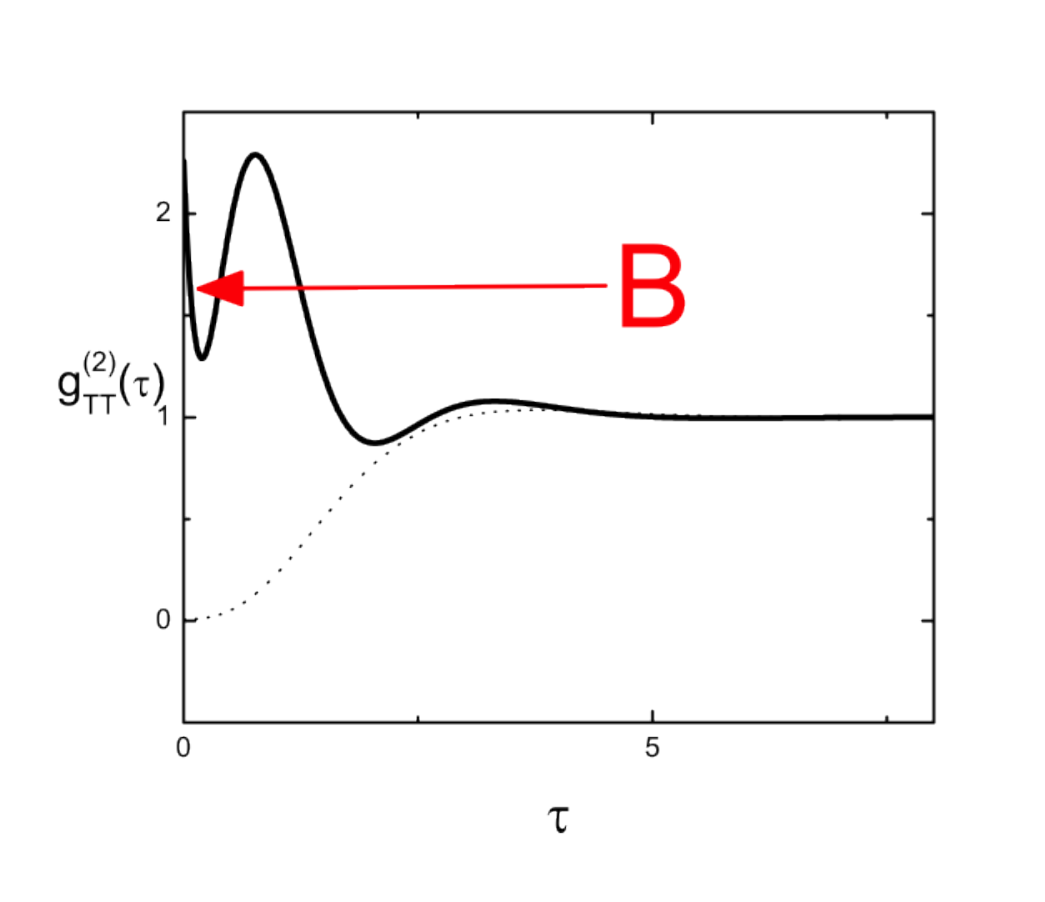}
    \caption{\emph{Bunching represented by \lq\lq
    B\rq\rq.}}\label{Bunchingpic}\label{ExampleB}
\end{figure}

\begin{figure}
\centering
         \includegraphics[width=\textwidth,totalheight=2.5in, keepaspectratio]{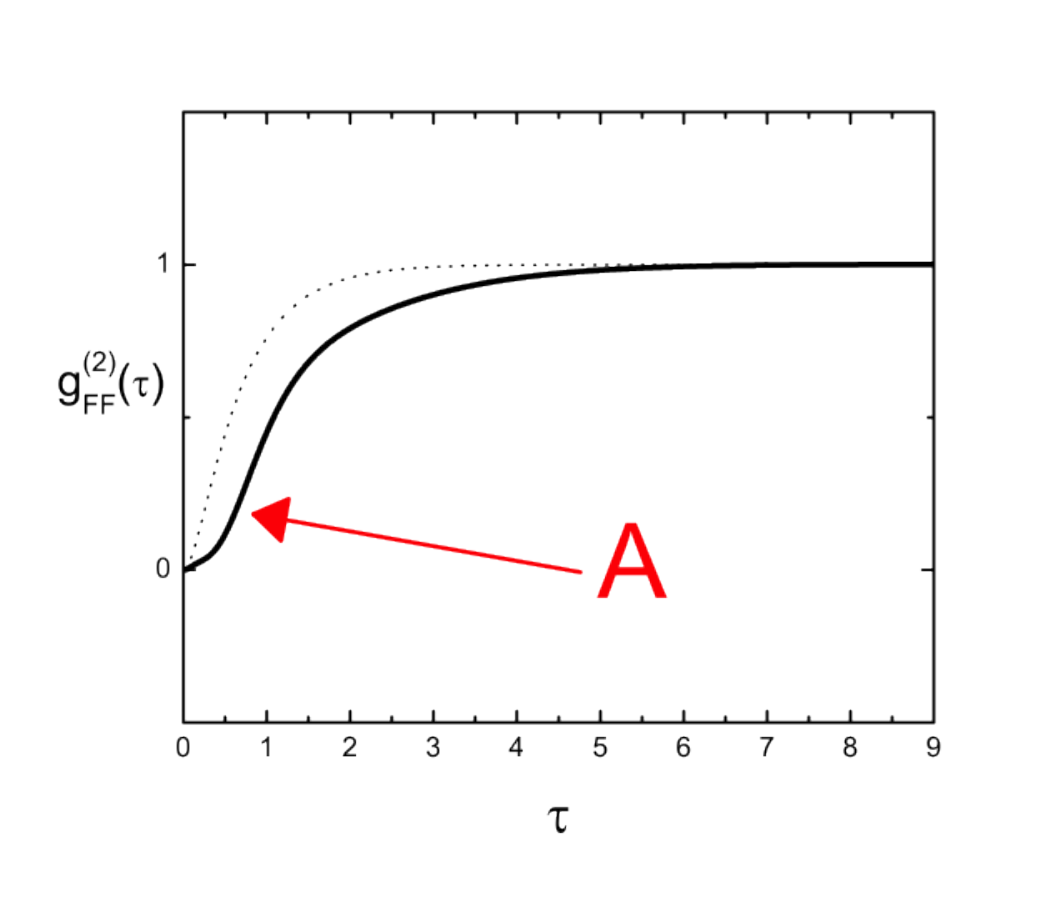}
    \caption{\emph{Anti-bunching represented by \lq\lq
    A\rq\rq.}}\label{Bunchingpic}\label{ExampleA}

\end{figure}

\clearpage

\begin{figure}
    \centering
         \includegraphics[width=\textwidth,totalheight=2in,
         keepaspectratio]{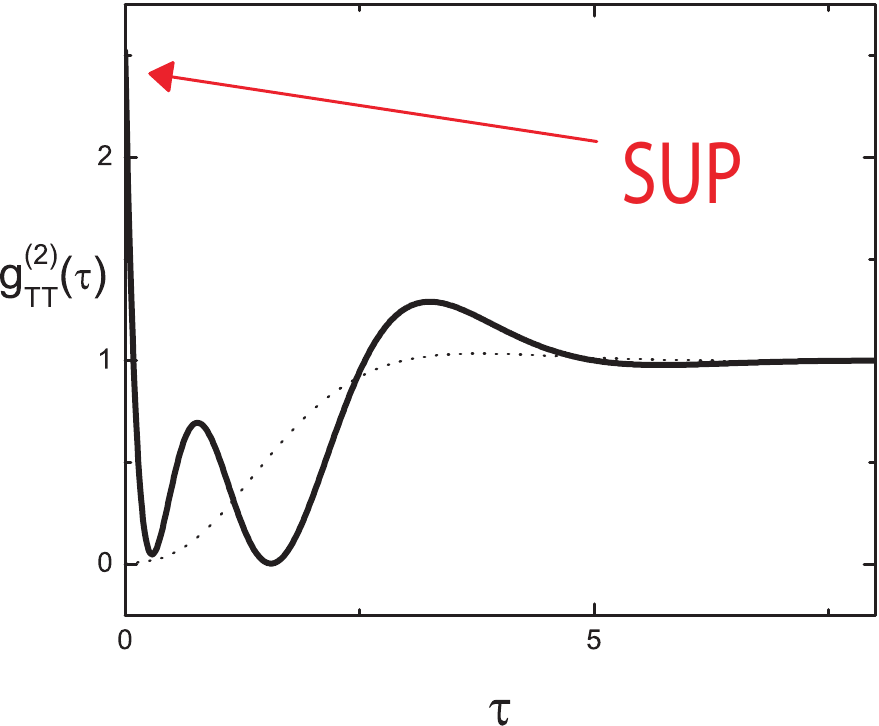}\label{ExampleSuperPoissonian}
    \caption{\emph{Example of a graph exhibiting Super-Poissonian statistics, denoted \lq\lq
    SUP\rq\rq}}
\end{figure}

The next inequality is derived from the fact that all classical
well behaved functions must obey the Schwartz inequality.
  Because we known
 that the general form of the Schwartz Inequality is

\footnotesize
\begin{eqnarray}\label{Schwartz1}
    \left(\int dx \int dy |f(x,y)g(x,y)|P(x,y)\right)^2  &\leq&
    \left(\int dx \int dy {f(x,y)}^2 P(x,y)\right) \nonumber \\
    &\times& \left(\int dx \int dy \quad {g(x,y)}^2 P(x,y)\right).
\end{eqnarray}
\normalsize

 We can absorb the probability into a
function of $x,y$ and write

 \footnotesize
\begin{eqnarray}\label{Schwartz11}
    \left(\int dx \int dy |\bar{f}(x,y)\bar{g}(x,y)|\right)^2  &\leq&
    \left(\int dx \int dy {\bar{f}(x,y)}^2 \right) \nonumber \\
    &\times& \left(\int dx \int dy \quad {\bar{g}(x,y)}^2 \right).
\end{eqnarray}
\normalsize By choosing $x = \bar{I}$, $y = \bar{I}_0$, $f(x,y) =
\bar{I}$,$\quad g(x,y) = 1$, and $P(x,y) = P(\bar{I}, t + \tau ;
\bar{I}_0, t)$ where $P(\bar{I}, t + \tau ; \bar{I}_0, t)$ is the
joint probability function that there is field intensity $\bar{I}$
at time $t +\tau$ and intensity $\bar{I}_0$ at time $t$, equation
\ref{Schwartz1} becomes
\begin{eqnarray}\label{Schwartz2}
    \left(\int d \bar{I} \quad \bar{I}P(\bar{I}, t +\tau)\right)^2 &\leq&
     \left( \int d \bar{I} \quad{\bar{I}}^2 P(\bar{I}, t+ \tau)\right) \times 1 \nonumber \\
    {\langle \bar{I} \rangle}^2 &\leq& \langle {\bar{I}}^2 \rangle \nonumber \\
    \frac{\langle {\bar{I}}^2 \rangle}{{\langle \bar{I} \rangle}^2} &\geq&
    1.
\end{eqnarray}

And now we express the Intensities in terms of the field as

\begin{eqnarray}\label{Schwartz3}
    \frac{\langle {E^*(t)}^2 {E(t)}^2 \rangle}{{E^{\ast}(t) E(t) \rangle}^2} &\geq& 1 \nonumber \\
    \frac{\langle {a^{\dag}(t)}^2 {a(t)}^2 \rangle}{{\langle {a^{\dag}}(t) a(t)\rangle}^2} &\geq&
    1.
\end{eqnarray}
Because the field is stationary, this can be written as
\begin{equation}\label{Schwartz4}
    \frac{\langle {a^{\dag}(t)}^2 {a(t)}^2 \rangle}{{\langle {a^{\dag}} a\rangle}^2} \geq
    1.
\end{equation}

Which is just the expression for $g^{(2)}(\tau)$ at time $t=0$, so
the final inequality is

\begin{equation}
g^{(2)}(0) \ge 1
\end{equation}

no non-negative probability distributions occur (refer to section
\ref{g2chpt} for explanation). If $g^{(2)}(0)
> 1$ our data is super-Poissonian, and if If $g^{(2)}(0)
< 1$ our data is sub-Poissonian.  They shall be referred to as
\lq\lq SUP\rq\rq\ and \lq\lq SUB\rq\rq.  An example of the
sub-Poissonian condition is shown in figure
\ref{ExampleSuperPoissonian}.  Please note specifically the
notation used in this section, as some people refer to both of
these violations as anti-bunching.  

The next set of equalities represent what shall be termed an
overshoot \lq\lq OS\rq\rq\ or an undershoot \lq\lq US\rq\rq\.
These are represented as

\begin{eqnarray}
|g^{(2)}(\tau) - 1| > |g^{(2)}(0) - 1|\\
|g^{(2)}(\tau) - 1| < |g^{(2)}(0) - 1|
\end{eqnarray}

respectively.  An example of each of these is shown (see figures
\ref{ExampleOvershoot}, \ref{ExampleUndershoot}).

\begin{figure}
    \centering
         \includegraphics[width=\textwidth,totalheight=2.5in,
         keepaspectratio]{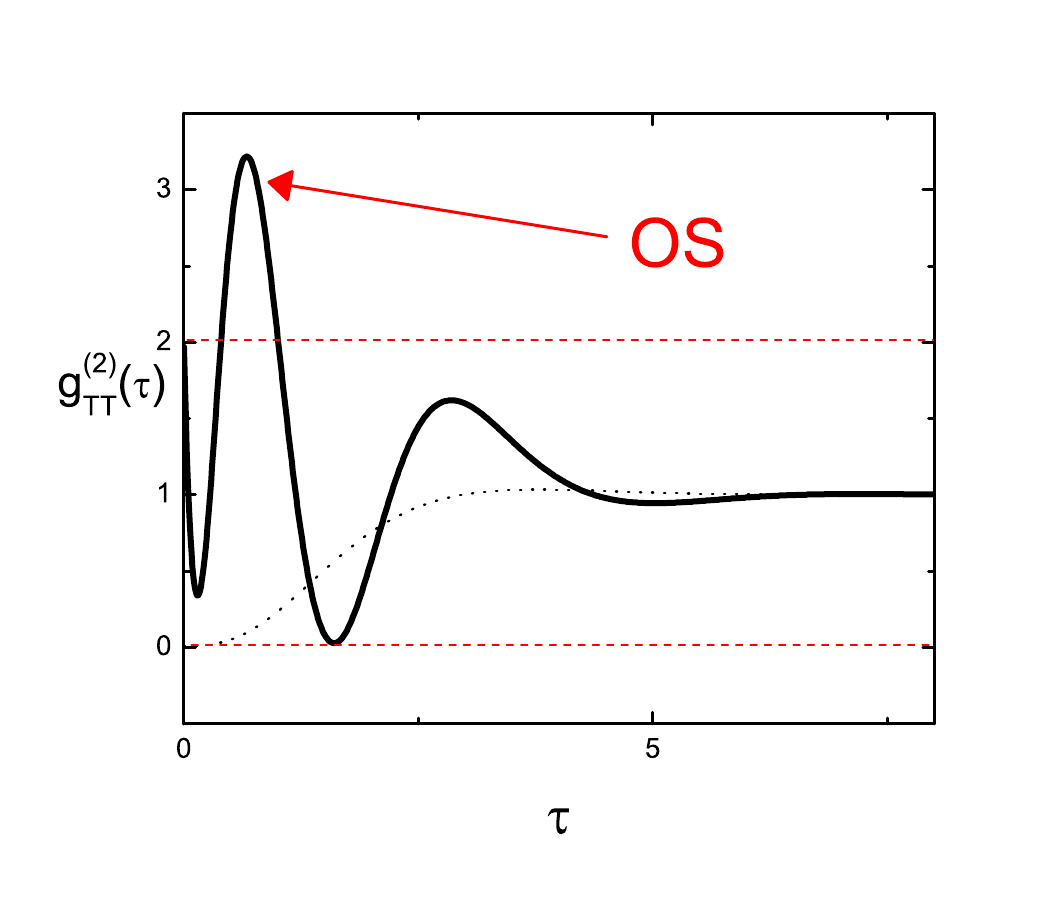}\label{ExampleOvershoot}
    \caption{\emph{Overshoot represented by \lq\lq OS\rq\rq}}
\end{figure}

\begin{figure}
    \centering
         \includegraphics[width=\textwidth,totalheight=2.5in,
         keepaspectratio]{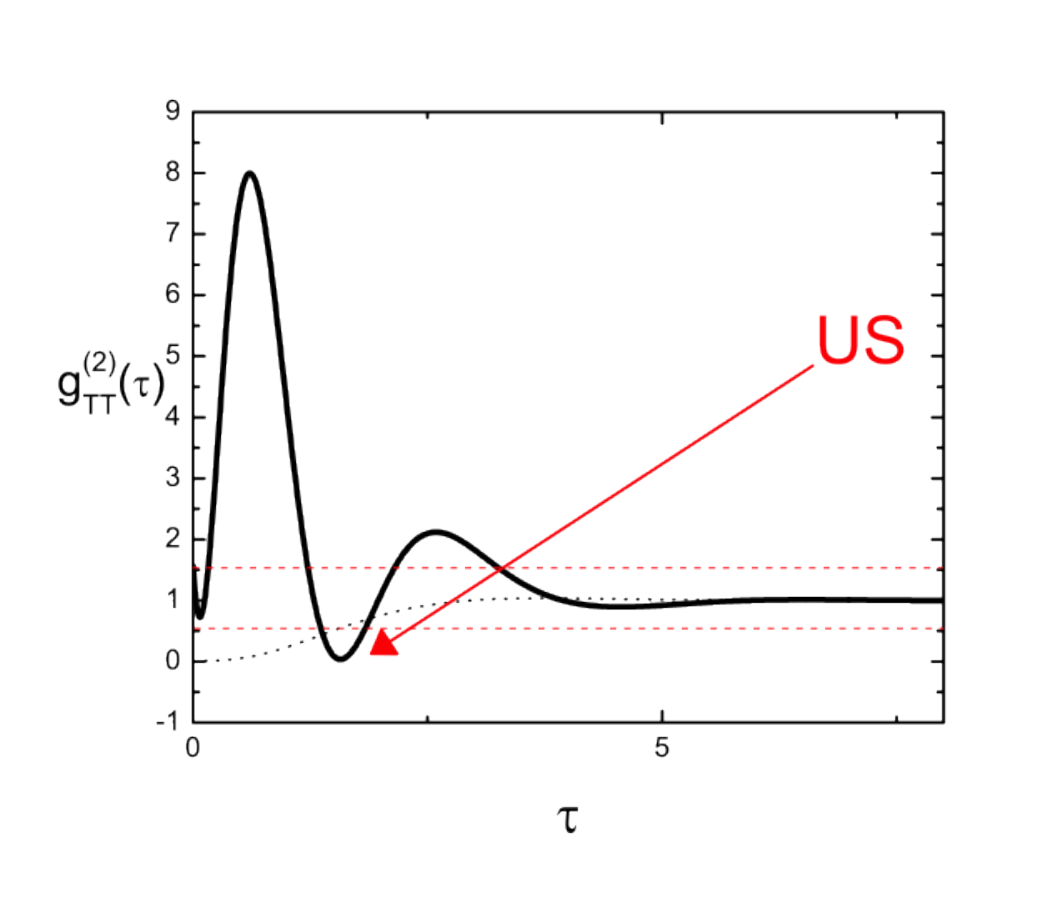}\label{ExampleUndershoot}
    \caption{\emph{Undershoot represented by \lq\lq US\rq\rq}}
\end{figure}
\clearpage

\subsection{Cross Correlations}
\setlength{\parindent}{0.25in}

There are only two inequalities that must be examined for the
cross-correlations.  They will be called  Cross-Violations, and
denoted as \lq\lq CV1\rq\rq and \lq\lq CV2\rq\rq.  A CV1 disobeys
the inequality

\begin{equation}
g^{(2)}_{TF,FT}(0) \le \sqrt{g^2_{TT}(0) g^2_{FF}(0)}
\end{equation}

However, because the case being examined is for one atom,
$g^2_{FF}(0)$ is always zero.  This is because of the fact that
$\sigma_-|e\rangle = |g\rangle$, and $\sigma_-|g\rangle$ is
impossible (refer to section \ref{TFFT}).  This simplifies the CV1
to

\begin{equation}
g^{(2)}_{TF,FT}(0) \le 0
\end{equation}

in which case there will always be a violation.  As for CV2,

\begin{equation}
g^2_{TF}-1 \le \sqrt{|g^2_{TT}-1||g^2_{FF}-1|}
\end{equation}

which again simplifies to

\begin{equation}
g^2_{TF}-1 \le \sqrt{g^2_{TT}-1}
\end{equation}

\clearpage
\section{Graphs for $g/\gamma = 1$, $\kappa/\gamma = 1.6$}
\setlength{\parindent}{0.25in}

Located on each graph will be a small table indicating which
non-classical behaviors are present.  If a behavior is not listed,
it is assumed to be classical.  The following is a reminder of
each non-classical abbreviation.

\begin{center}

B - Bunching\\
A - Anti-bunching\\
SUB - Sub-Poissonian probability distribution\\
SUP - Super-Poissonian probability distribution\\
OS - Overshoot\\
US - Undershoot\\

CV1 - Cross-violation 1\\
CV2 - Cross-violation 2\\
\end{center}

Note also that some of the graphs are not smooth lines, but
instead have "wiggles".  What is being seen are the beat
frequencies.  In the dressed-state picture, each next highest
level can be considered an un-coupled three-level system.  When
these interact, we see the beat frequencies.

\clearpage

\section{Inequalities and Non-Classical Behaviors}
\setlength{\parindent}{0.25in}

As described in Section \ref{Squeezing}, a violation of the
classical behaviors of $h_\theta$ are a sign of squeezing (refer
to section 7.2.4). Again, the results will be divided into two
categories - the transmissions and fluorescence, and the cross
correlations.

\subsection{Transmission and Fluorescence}
\setlength{\parindent}{0.25in}

There are only two possible violations to consider
\cite{{ElliotRice},{ElliotJoeRice}}.  The violations will be
denoted as \lq\lq S1\rq\rq and \lq\lq S2\rq\rq because they both
signify squeezing. They are defined respectively as

\begin{eqnarray}
0 \le h_\theta(0) - 1 \le 1\\
|h_\theta(\tau) -1 | \le |h_\theta(0)-1|\le 1
\end{eqnarray}

An S1 violation will therefore occur any time the value of
$h_\theta(0)$ is not between 1 and 2.  Note that because the
fluorescence condition has an initial value of zero, it will
always have an S1 violation.  An S2 violation will occur any time
the graph dips below zero or rises above two.  Furthermore, an S2
violation can occur between a more narrow range of values,
dependant upon the initial value, as analogous to
overhoot/undershoot violations for $g^{(2)}(\tau)$. Examples of S1
and S2 violations are shown below

\begin{figure}
    \centering
         \includegraphics[width=\textwidth,totalheight=2in,
         keepaspectratio]{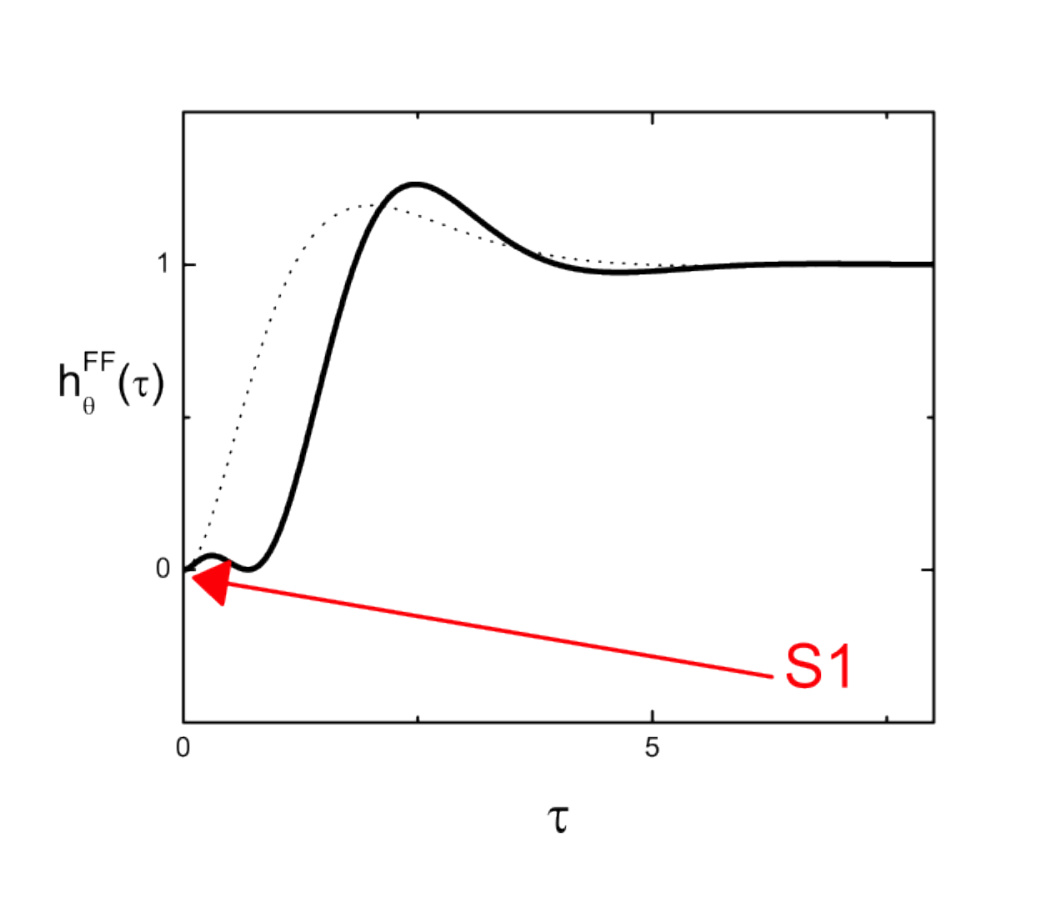}
    \caption{\emph{Example of a graph exhibiting a S1
    violation.}}\label{S1ExampleS1}
\end{figure}

\begin{figure}
    \centering
         \includegraphics[width=\textwidth,totalheight=2in,
         keepaspectratio]{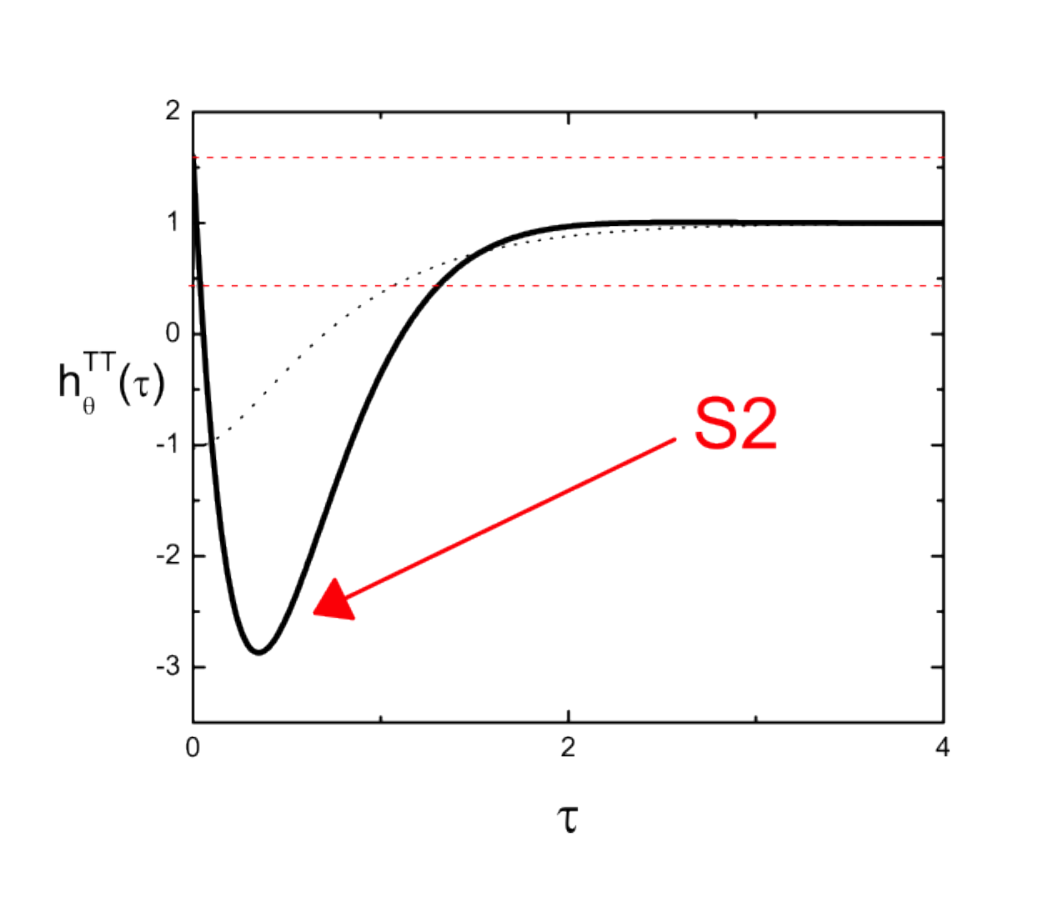}
    \caption{\emph{Example of a graph exhibiting a S2 violation.}\label{S2ExampleS2}}
\end{figure}

Finally we consider violations of classical inequalities for the cross correlations. These are also readily observed

\begin{figure}
    \centering
         \includegraphics[width=\textwidth,totalheight=2in,
         keepaspectratio]{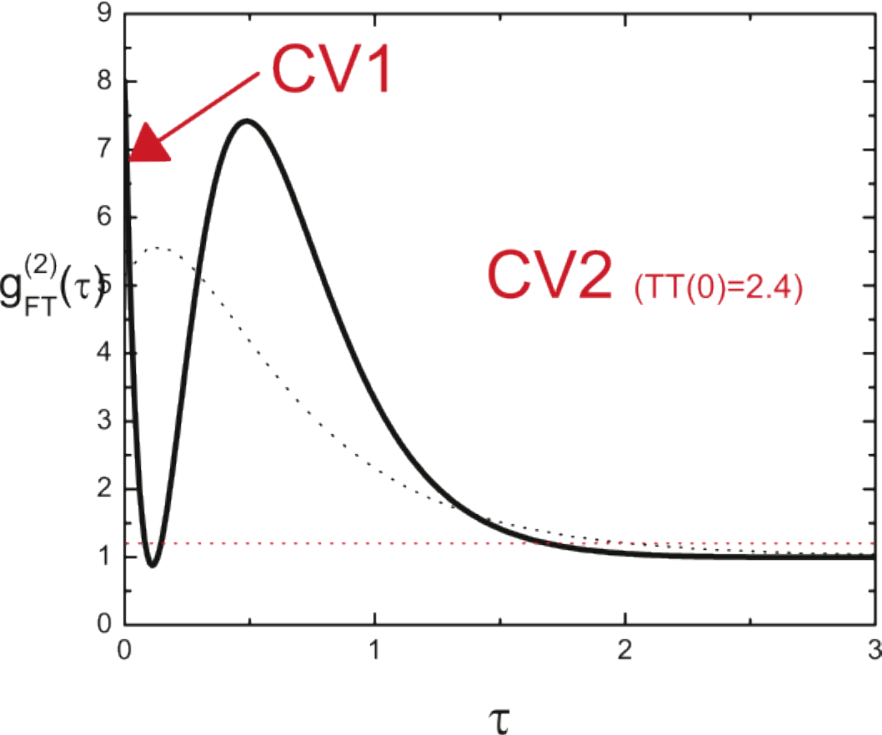}
    \caption{\emph{Example of a graph exhibiting CV1 and CV2
    violations.}}
\end{figure}

We now present tables that examine which nonclassical effects happen for which parameters

\begin{figure}
    \centering
         \includegraphics[width=\textwidth,totalheight=6in,
         keepaspectratio]{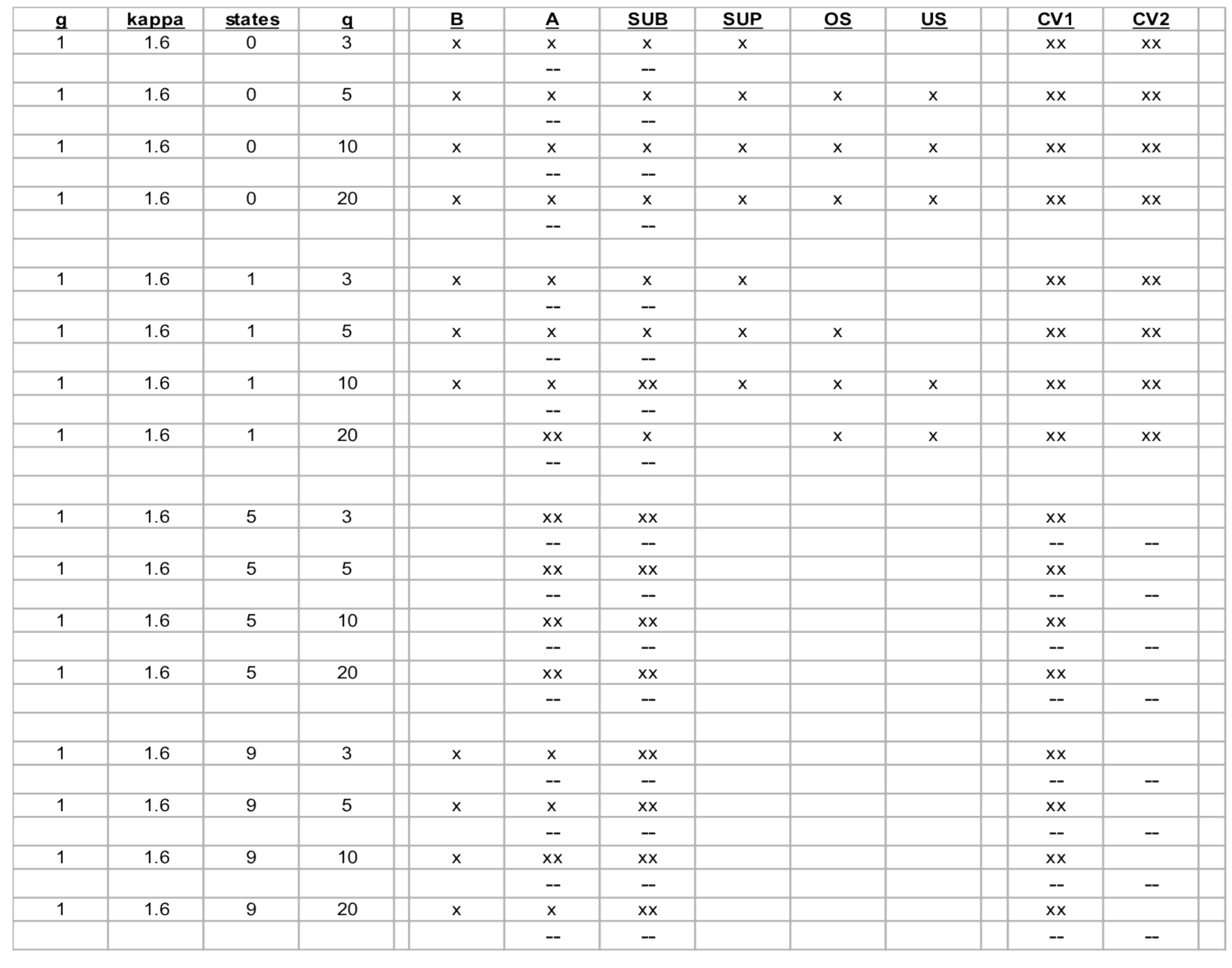}
     \caption{\emph{Nonclassical effects in $g^{(2)}(\tau)$.}}
\end{figure}

\begin{figure}
    \centering
         \includegraphics[width=\textwidth,totalheight=6in,
         keepaspectratio]{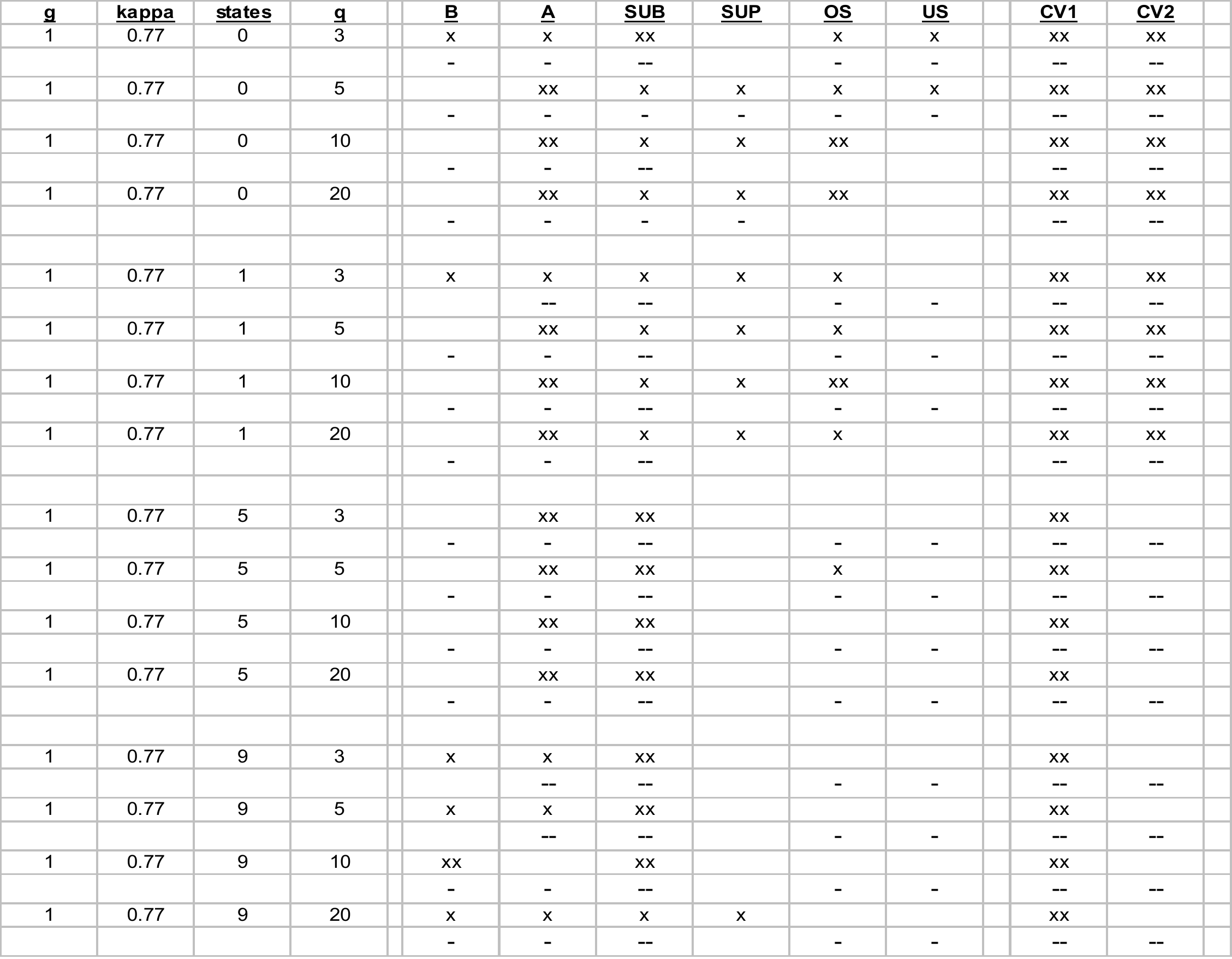}
    \caption{\emph{Nonclassical effects in $g^{(2)}(\tau)$.}}
\end{figure}

\begin{figure}
    \centering
         \includegraphics[width=\textwidth,totalheight=6in,
         keepaspectratio]{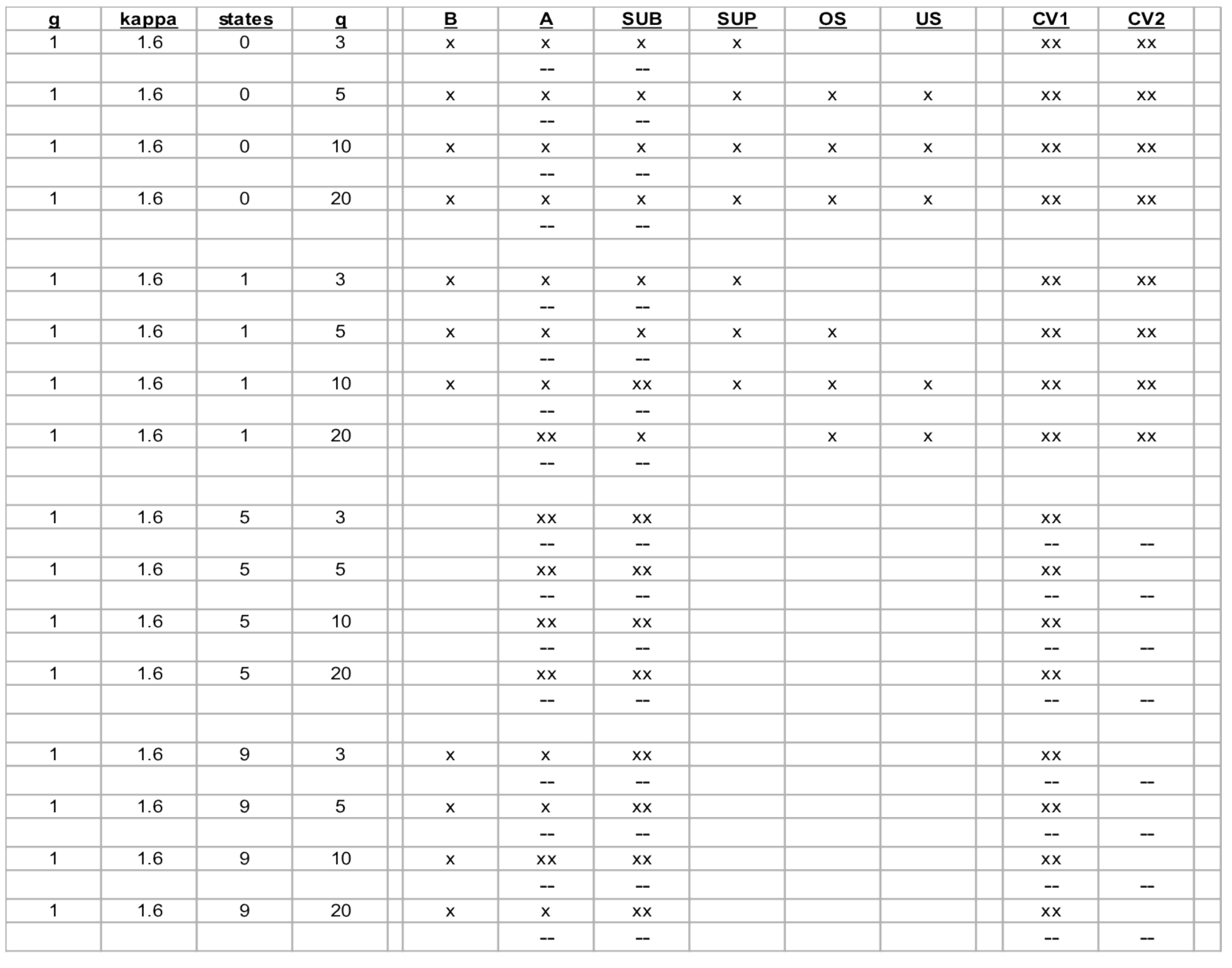}
    \caption{\emph{Nonclassical effects in $g^{(2)}(\tau)$.}}
\end{figure}

\begin{figure}
    \centering
         \includegraphics[width=\textwidth,totalheight=6in,
         keepaspectratio]{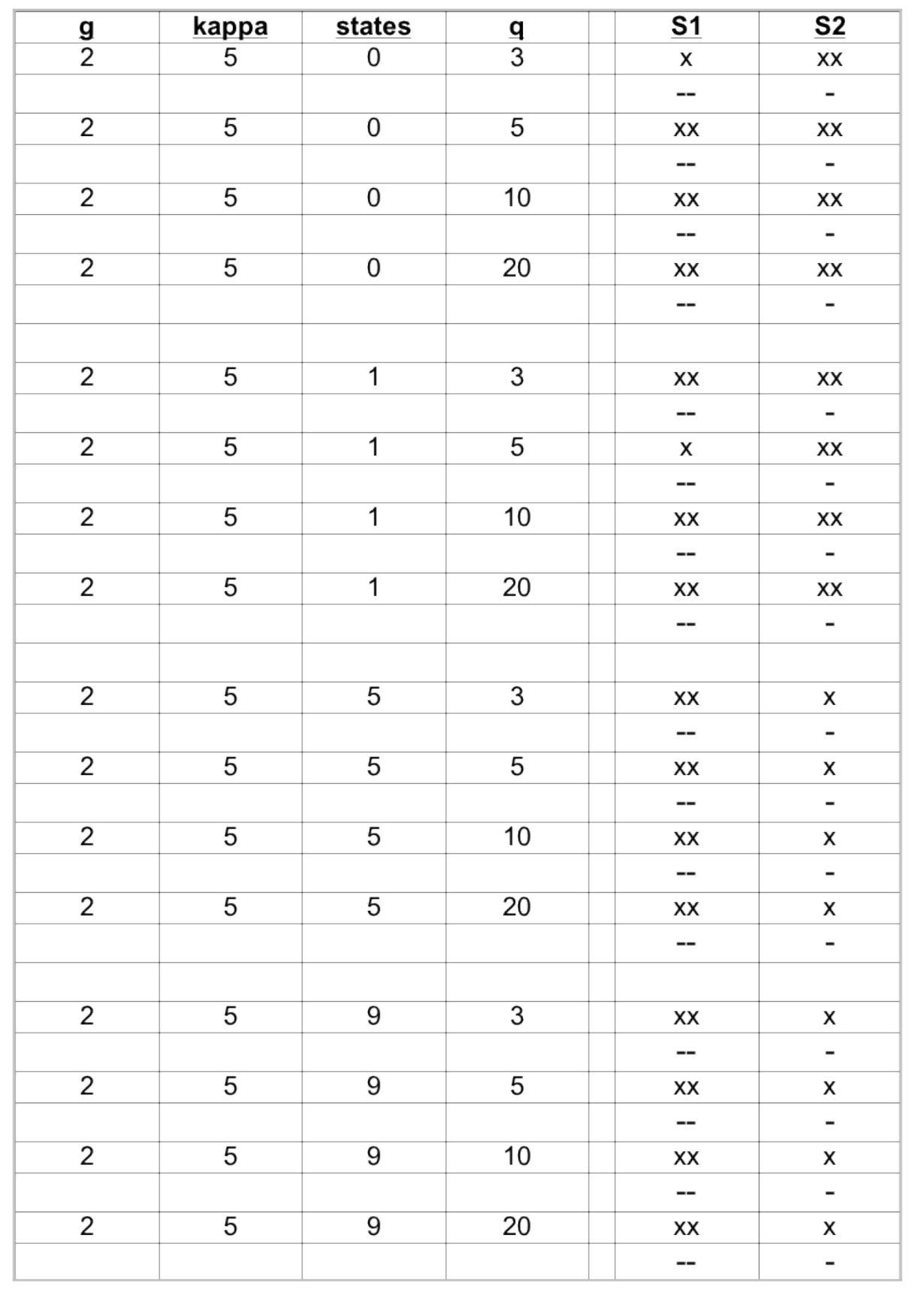}
    \caption{\emph{Nonclassical effects in $h_{\theta}(\tau)$}}
\end{figure}

\section{Conclusions}
We have investigated intensity-intensity and field-intensity correlations in a cavity QED system with an internal potential with a periodity of $\lambda/1$ where $\lambda$
is the simultaneous wavelength of the atomic transition and cavity mode. When both the atom and cavity are off
resonance it was found that the anti-bunching in all the cases
disappeared the more off resonance we went. It was found that both
the photon statistics and the wave-particle correlation functions
are quite sensitive to center-of-mass wave function. This can be eased by choosing atomic and cavity detunings equal and opposite (in units of their respective linewidths). Here nonclassical 
behavior is not reduced drastically We saw that
an increase in the width of the Gaussian in both the SHO and
Mahtieu cases washed away non-classicalities with the Mathieu case
experiencing more rapid washing out with increase in Gaussian
width. In all cases investigated, the correlation functions
$g^{(2)}_{FF}$ and $h^{FF}_{\theta}$ appear to be sensitive only
to the Mathieu and SHO population distributions for large values
of the Gaussian width. The intensity-field fluctuations are not as sensitive to detunings.

\end{document}